\documentstyle[aps]{revtex}
\begin{document}
\author{Changrim Ahn, Kong-Ju-Bock Lee}
\address{Department of Physics, Ewha Womans University,
Seoul, 120-750, Korea}
\author{Soonkeon Nam}
\address{Department of Physics, Kyung Hee University,
Seoul, 130-701, Korea}
\title{Nonrelativistic Factorizable Scattering Theory
of Multicomponent Calogero-Sutherland Model}
\maketitle
\begin{minipage}{\textwidth}
\begin{quotation}
\begin{abstract}

We relate two integrable models in (1+1) dimensions, namely,
multicomponent Calogero-Sutherland
model with particles and antiparticles interacting via the
hyperbolic potential and
the nonrelativistic factorizable $S$-matrix theory with
$SU(N)$-invariance.
We find complete solutions of the Yang-Baxter equations without
implementing the crossing symmetry,
and one of them is identified with
the scattering amplitudes derived from
the Schr\"{o}dinger equation of the Calogero-Sutherland model.
This particular solution is of interest in that it cannot be
obtained
as a nonrelativistic limit of any known relativistic solutions
of the $SU(N)$-invariant Yang-Baxter equations.

\bigskip
\end{abstract}

\end{quotation}
\end{minipage}

Recently, there has been a great interest in integrable quantum
systems with long-range interactions. Of these the
Calogero-Sutherland model (CSM) \cite{Calogero,Sutherland},
which is an $n$-body non-relativistic quantum mechanical system
with long-range two-body potentials, has been considered in
connection with integrable spin-chains with long-range
interactions \cite{Haldane,Shastry},
random matrix theory \cite{simons},
and fractional statistics \cite{fracstat}.
Furthermore it was shown in \cite{ZamZam},
the $S$-matrix of the model with hyperbolic potentials
is the same as the non-relativistic limit of the sine-Gordon
soliton $S$-matrix.
That is, the $S$-matrix of CSM is a $O(2)$-symmetric
solution of the Yang-Baxter equations, together with
unitarity condition, but without the crossing symmetry.
This connection was also observed for the boundary sine-Gordon
equation in its relation to $BC_n$ type CSM \cite{Skorik}.

In this letter, we establish a connection between
the multicomponent CSM,
the $n$-body quantum mechanical
system of colored particles and antiparticles interacting via
integrable long-range potential of hyperbolic-type and
nonrelativistic factorizable $S$-matrix theory with
$SU(N)$-invariance.
First we derive the scattering amplitudes from the
eigenfunctions of the CSM Hamiltonian.
To relate this to the $S$-matrix theory, we obtain complete
solutions of the $SU(N)$-invariant Yang-Baxter equations.
In relativistic scattering theory, this model has been solved
completely in \cite{Berg}.
Without implementing the crossing symmetry we find a new class of
solutions and, interestingly, one of these corresponds to the
scattering amplitudes of the CSM.
This means that the multicomponent CSM cannot be obtained as a
nonrelativistic limit of any relativistic systems, contrary to
the the sine-Gordon model ($O(2)$ case) mentioned above.

We begin with the multicomponent CSM where the Hamiltonian is
given by
\begin{equation}
H = - \sum_i^{n} \frac{\partial^2}{\partial x_i} + \sum_{i<j}
\frac{\lambda (\lambda+P_{ij})}{\sinh ^2 (x_i - x_j)}
\end{equation}
where $P_{ij}$ is the exchange operator of the colors of
(anti)particles at $x_i$ and $x_j$.
This model has been shown to be integrable by several authors
\cite{Poly,Ha}.
Since $P_{ij}^2 = I$, we can define the eigenstates of $P_{ij}$
as
$|\pm \rangle = \frac{1}{\sqrt{2}} \left( |\sigma_{i}\sigma_{j}
\rangle\pm |\sigma_{j}\sigma_{i}\rangle\right)$
such that $P_{ij} |\pm\rangle = \pm |\pm\rangle $.

To obtain the two-body $S$-matrix, we consider the scattering
eigenstates of the Schr\"{o}dinger equation
$\left[ -\frac{d^2}{dx^2} + \frac{\lambda(\lambda+1)}{\sinh^2 x}\right]
\psi_k(x) = k^2 \psi_k(x)$.
Due to the underlying $SU(1,1)$ structure of the scattering
problem \cite{gursey}, $\psi_k(x)$ is proportional to
$(\sinh x)^{\lambda +1}\ _{2}F_{1}
((\lambda +1+ik)/2,(\lambda +1-ik)/2,\lambda +3/2;-\sinh^2 x)$,
where $\ _{2}F_{1}$ is the hypergeometric function.
The asymptotic states for $x\rightarrow \infty$ is
\begin{equation}
\psi_k(x) \rightarrow C\left(e^{ikx}
\frac{\Gamma(ik)\Gamma(2\lambda +2)}{\Gamma(\lambda+1+ik)\Gamma
(\lambda+1)}
+e^{-ikx}\frac{\Gamma(-ik)\Gamma(2\lambda +2)}{\Gamma(\lambda+1-ik)
\Gamma(\lambda+1)}\right)
\end{equation}
the two-body scattering matrices are
\begin{equation}
S^+ (k) = \frac{\Gamma(ik) \Gamma(1+\lambda -ik)}{\Gamma(-ik)
\Gamma(1+\lambda+ik)}, \qquad
S^- (k) = \frac{\Gamma(ik) \Gamma(\lambda -ik)}{\Gamma(-ik)
\Gamma(\lambda+ik)},
\end{equation}
for $P_{ij}=\pm 1$, respectively\cite{Perelomov}.

Now returning to $\vert\sigma_{i}\sigma_{j}\rangle$ basis,
it is straightforward to
obtain particle-particle scattering amplitudes,
\begin{equation}
S^{\sigma_j\sigma_i}_{\sigma_i\sigma_j}\equiv u_1 =
\frac{1}{2}(S^{+}+S^{-}),
\qquad S^{\sigma_i\sigma_j}_{\sigma_i\sigma_j}\equiv u_2 =
\frac{1}{2}(S^{+}-S^{-}),
\end{equation}
for $\sigma_i\neq\sigma_j$. Note that for $\sigma_i=\sigma_j$,
$S^{\sigma_i\sigma_i}_{\sigma_i\sigma_i} = u_1+u_2$.
Similarly, the scattering amplitudes of particles and
antiparticles of different colors,
$\sigma_i \neq \bar\sigma_j$
become
\begin{equation}
S^{\sigma_i\bar\sigma_j}_{\sigma_i\bar\sigma_j}\equiv r_1 =
\frac{1}{2}(S^{+}+S^{-}), \qquad
S^{\bar{\sigma}_j\sigma_i}_{\sigma_i\bar\sigma_j}\equiv t_1 =
\frac{1}{2}(S^{+}-S^{-}),
\end{equation}
where $\bar\sigma_j$ stands for the color of an antiparticle.
(See Fig.1 for schematic definitions of the amplitudes.)

\begin{picture}(600,180)(0,0)
\put(70,110){\line(-1,1){40}}
\put(30,110){\line(1,1){40}}
\put(150,110){\line(-1,1){40}}
\put(110,110){\line(1,1){40}}
\put(230,110){\line(-1,1){40}}
\put(190,110){\line(1,1){40}}
\put(70,20){\line(-1,1){40}}
\put(30,20){\line(1,1){40}}
\put(150,20){\line(-1,1){40}}
\put(110,20){\line(1,1){40}}
\put(230,20){\line(-1,1){40}}
\put(190,20){\line(1,1){40}}
\put(20,102){$\sigma_{i}$}
\put(70,102){$\sigma_{j}$}
\put(20,155){$\sigma_{j}$}
\put(70,155){$\sigma_{i}$}
\put(45,90){$u_1$}
\put(100,102){$\sigma_{i}$}
\put(150,102){$\sigma_{j}$}
\put(100,155){$\sigma_{i}$}
\put(150,155){$\sigma_{j}$}
\put(125,90){$u_2$}
\put(180,102){$\sigma_{i}$}
\put(230,102){${\overline\sigma}_{j}$}
\put(180,155){$\sigma_{i}$}
\put(230,155){${\overline\sigma}_{j}$}
\put(205,90){$r_1$}
\put(20,12){$\sigma_{i}$}
\put(70,12){${\overline\sigma}_{i}$}
\put(20,65){$\sigma_{j}$}
\put(70,65){${\overline\sigma}_{j}$}
\put(45,0){$r_2$}
\put(100,12){$\sigma_{i}$}
\put(150,12){${\overline\sigma}_{j}$}
\put(100,65){${\overline\sigma}_{j}$}
\put(150,65){$\sigma_{i}$}
\put(125,0){$t_1$}
\put(180,12){$\sigma_{i}$}
\put(230,12){${\overline\sigma}_{i}$}
\put(180,65){${\overline\sigma}_{j}$}
\put(230,65){$\sigma_{j}$}
\put(205,0){$t_2$}
\put(10,110){\vector(0,1){30}}
\put(0,145){time}
\end{picture}
\vskip .3cm
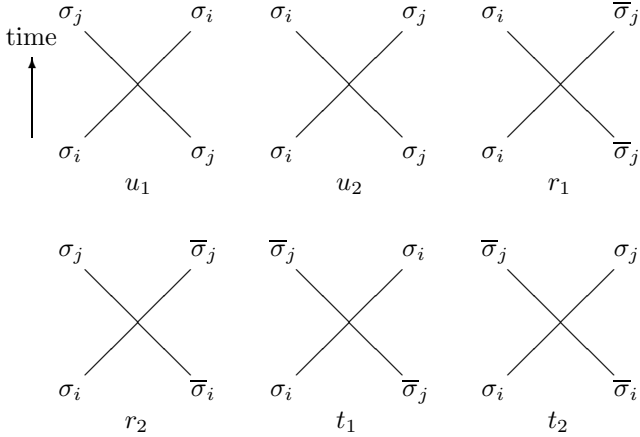
{\centerline{\small Fig.1: Scattering amplitudes of particles and
antiparticles with colors}}
\vskip .5cm

The scattering amplitudes of the same color have to be dealt
with some care.  The most general eigenstates of $P_{ij}$ are
now of the form
$|\pm\rangle=\frac{1}{\sqrt{2}}(|A\rangle \pm |B\rangle)$,
where
$|A\rangle = {\cal N}\sum^{N}_{i=1}a_i |\sigma_i\bar\sigma_i\rangle$
and
$|B\rangle = {\cal N}\sum^{N}_{i=1}a_i |\bar\sigma_i\sigma_i\rangle$
for some coefficients $a_i$'s and ${\cal N}=
(\sum^{N}_{i=1}a_i^2)^{-1/2}$.
The scattering amplitudes for $|\pm\rangle$ are $S^{\pm}$ as
before and if
we express these in the $|\sigma_i\bar\sigma_i\rangle$ basis we find
\begin{equation}
\frac{1}{2}(S^+ + S^-)=r_1 + M r_2,\qquad
\frac{1}{2}(S^+ - S^-)=t_1 + M t_2,
\qquad {\rm where}\qquad M={\cal N}^2\sum_{i,j}^{N} a_i a_j.
\end{equation}
Comparing Eq.(5) with (6), we find $r_2=t_2=0$.
With $S^{\pm}$ given in Eq.(3), we obtain the scattering
amplitudes of the $SU(N)$-invariant CSM as follows:
\begin{equation}
u_1=t_1=\frac{-ik}{\lambda+ik}\cdot\frac{\Gamma(ik)\Gamma
(\lambda-ik)}{\Gamma(-ik)\Gamma(\lambda+ik)}, \quad
u_2=r_1=\frac{\lambda}{\lambda+ik}\cdot\frac{\Gamma(ik)\Gamma
(\lambda-ik)}{\Gamma(-ik)\Gamma(\lambda+ik)}, \quad r_2=t_2=0.
\end{equation}

Since we have derived the scattering amplitudes of the CSM which
is integrable, it is reasonable to study the problem in the
context of the factorizable scattering theory where the
integrability requires the Yang-Baxter equations.
As a candidate, we consider the  Yang-Baxter equations with
$SU(N)$-invariance because the particles and antiparticles
of the CSM carry the $SU(N)$-color quantum numbers.
For the nonrelativistic system where CPT-invariance does
not hold anymore, one needs not implement the crossing symmetry.

The Yang-Baxter equations and unitarity condition give 15
equations for the amplitudes
which are the same as the relativistic case,
Berg et. al. \cite{Berg}.  Seven of these include neither
$r_2$ nor $t_2$;
\begin{eqnarray}
u_1(\theta) u_1(-\theta) &+& u_2(\theta)u_2(-\theta)=1, \\ \nonumber
u_1(\theta) u_2(-\theta) &+& u_2(\theta)u_1(-\theta)=0, \\ \nonumber
t_1(\theta) t_1(-\theta) &+& r_1(\theta)r_1(-\theta)=1, \\ \nonumber
t_1(\theta) r_1(-\theta) &+& r_1(\theta)t_1(-\theta)=0, \\ \nonumber
u_1(\theta)t_1(\theta+\theta')r_1(\theta') &
=& t_1(\theta)u_1(\theta+\theta')r_1(\theta'),\\ \nonumber
u_2(\theta)t_1(\theta+\theta')r_1(\theta') &=& r_1(\theta)r_1
(\theta+\theta')t_1(\theta')+
t_1(\theta)u_2(\theta+\theta')r_1(\theta'),\\  \nonumber
u_2(\theta)u_1(\theta+\theta')u_2(\theta') &=& u_1(\theta)u_2
(\theta+\theta')u_2(\theta')+
u_2(\theta)u_2(\theta+\theta')u_1(\theta'), \nonumber
\end{eqnarray}
where $\theta$ is the spectral parameter.
We see that this has the ``minimal'' solution of
\begin{equation}
u_1(\theta)= t_1(\theta)=\frac{-\theta}{\gamma+\theta}\cdot
\frac{\Gamma(\theta)\Gamma(\gamma-\theta)}{\Gamma(-\theta)
\Gamma(\gamma+\theta)},\qquad
u_2(\theta)= r_1(\theta)=\frac{\gamma}{\gamma+\theta}\cdot
\frac{\Gamma(\theta)\Gamma(\gamma-\theta)} {\Gamma(-\theta)
\Gamma(\gamma+\theta)},
\end{equation}
where $\gamma$ is the arbitrary parameter.
The rest of the equations are trivially
satisfied with $r_2=t_2=0$. Replacing $\theta$ with $ik$ and
interpreting the parameter $\gamma$ as the coefficient
of the hyperbolic interaction $\lambda$,
these scattering amplitudes are identical
to those of the multicomponent CSM.

We compute all other possible solutions which are
listed in Table 1.
It turns out that the classes I-VI are exactly the nonrelativistic
limits of the six classes of solutions that Berg et. al. \cite{Berg},
i.e., imposing the crossing
symmetry of $u_1(\theta)=t_1(i\pi-\theta),\ \ u_2(\theta)=
t_2(i\pi-\theta)$,
and $r_1(\theta)=r_2(i\pi-\theta)$, which fixes the parameter
$\gamma$ to a certain value for each classes.
The class VII corresponds to the multicomponent CSM.
The rest are new solutions which do not have relativistic analogs.

In Table 1, the functions $U(\theta),R(\theta)$, and $T(\theta)$
satisfy the unitarity condition $U(\theta)U(-\theta)=1$ etc.,
which can fix these functions with the minimality condition.
In the class X, functions $t_1,r_1$
satisfy $t_1(\theta)t_1(-\theta)+r_1(\theta)r_1(-\theta)=1$.

In our model, we have considered the case where the potentials
between two particles irrespective of the species are all
$1/\sinh^2 x$-type and shown that this corresponds to
$SU(N)$-invariant factorizable scattering theory.
We want to point out that there exists a more general
exactly solvable potential (P\"oschl-Teller)\cite{gursey}
which contains both $1/\sinh^2 x$ and $1/\cosh^2 x$\cite{romer}.
In ref.\cite{ZamZam}, it has been shown that
the $O(2)$-invariant scattering theory (the sine-Gordon
model) is related to the nonrelativistic Hamiltonian system
where the $1/\sinh^2 x$ potential is for
(anti)particle-(anti)particle scattering and the $1/\cosh^2 x$ for
particle-antiparticle.
It would be interesting to consider the case where
particle-antiparticle scattering potential is different from
particle-particle potential where particles carry color charges.
We would like to emphasize the approach to the multicomponent
CSM and the generalized Haldane-Shastry model
\cite{Haldane,Shastry,Kawa} based on
the factorizable $S$-matrix theory can be fruitful.
We hope our approach can be generalized to other integrable
Hamiltonian systems.

We would like to thank T.Deguchi, P.Fendley, N.Kawakami, C.Lee,
V.Pasquier, C.Rim, R.Sasaki for helpful discussion.
The work of CA and KL is supported in part by
KOSEF-941-0200-003-2
and by BSRI-94-2427(CA) and BSRI-94-2428(KL),
Ministry of Education, 1994.
The work of SN is by BSRI-94-2442 and Kyung Hee Univ. fund.
We also acknowledge the partial support of CTP/SNU.


\vskip .5cm
\begin{center}
\begin{tabular}{|c|c|c|c|c|c|c|}\hline

Class& $u_1(\theta)$ & $u_2(\theta)$ & $r_1(\theta)$ & $r_2(\theta)$
& $t_1(\theta)$ & $t_2(\theta)$ \\ \hline
I& ${\theta\over{\gamma\pm\theta}}U(\theta)$ &${\gamma\over{\theta}}
u_1(\theta)$  &0 &0 &$T(\theta)$&0 \\
II& ${\theta\over{\gamma\pm\theta}}U(\theta)$ &
${\gamma\over{\theta}}u_1(\theta)$ & 0 &0 &
$T(\theta)$ & $-{\gamma\over{{\gamma N\over{2}}+\theta}}
t_1(\theta)$ \\
III& $t_1(\theta)$ & $r_1(\theta)$ & ${\gamma\over{\theta}}
t_1(\theta)$&
${\gamma\over{\gamma(1-N)-\theta}}t_1(\theta)$&${\theta
\over{\gamma+\theta}}U(\theta)$&$r_2(\theta)$ \\
IV& $-t_1(\theta)$ & $r_1(\theta)$ & $-{\gamma\over{\theta}}
t_1(\theta)$&
${\gamma\over{-\gamma(1-N)-\theta}}t_1(\theta)$&${\theta\over
{\gamma-\theta}}U(\theta)
$&$r_2(\theta)$ \\
V& 0&$r_1(\theta)$ &$R(\theta)$& $r_1(\theta)$ &0&$r_1(\theta)$ \\
VI&0&$e^{\gamma\theta} r_1(\theta)$&$R(\theta)$&
$-{N(e^{2\gamma\theta}-1)\over{N^2 e^{2\gamma\theta}-1}}r_1(\theta)$
&0&$N^{-1}e^{-\gamma\theta}r_2(\theta)$ \\
VII&$t_1(\theta)$ & $r_1(\theta)$&$-{\gamma\over{\theta}}
t_1(\theta)$&0& ${-\theta\over{\gamma+\theta}}U(\theta)$& 0 \\
VIII&${\theta\over{\gamma\pm\theta}}U(\theta)$ &${\gamma
\over{\theta}}u_1(\theta)$
&$R(\theta)$ &0 &0&0 \\
IX&0&$U(\theta)$&$R(\theta)$&0&0&$\gamma\left[U(\theta)-U(-\theta)
{R(\theta)\over{R(-\theta)}}\right]$ \\
X&0&$U(\theta)$&$r_1(\theta)$&0&$\gamma\left[U(\theta)-U(-\theta)
{r_1(\theta)\over{r_1(-\theta)}}\right]$
&0\quad{\rm or}\quad$-{N\over{2}} t_1(\theta)$ \\
\hline
\end{tabular}
\end{center}
\centerline{\small Table 1: Complete solutions of nonrelativistic
$SU(N)$ invariant Yang-Baxter equations}
\vskip .5cm
\end {document}